\documentclass[aps,twocolumn,floats,pra,superscriptaddress,longbibliography]{revtex4-2}
\usepackage[colorlinks,linkcolor=black,citecolor=blue,urlcolor=black]{hyperref}
\usepackage{amsmath,amsfonts,amssymb,graphicx,bm,color,appendix}
\begin{document}

\author{G. Spada}
\affiliation{Dipartimento di Fisica, Universit\`a di Trento and CNR-INO BEC Center, 38123 Povo, Trento, Italy}
\author{S. Pilati}
\affiliation{School of Science and Technology, Physics Division, Universit\`a di Camerino, 62032 Camerino, Italy}
\affiliation{INFN-Sezione di Perugia, 06123 Perugia, Italy}
\author{S. Giorgini}
\affiliation{Dipartimento di Fisica, Universit\`a di Trento and CNR-INO BEC Center, 38123 Povo, Trento, Italy}

\title{Thermodynamics of a dilute Bose gas: A path-integral Monte Carlo study}
\begin{abstract}
We present precise path-integral Monte Carlo results for the thermodynamics of a homogeneous dilute Bose gas. Pressure and energy are calculated as a function of temperature both below and above the Bose-Einstein transition. Specifically, we address interaction effects, focusing on deviations from the ideal gas law in the thermodynamic limit. We also calculate the isothermal compressibility and the contact parameter, which provide a clear signature of the role played by interactions. In particular, we obtain indications of a discontinuity of the compressibility at the transition point. To gain physical insight, numerical results are systematically compared with the predictions of first-order Hartree-Fock and second-order Popov theories, both giving an approximate description of the gas thermodynamics. The comparison shows the extension of the critical region around the transition point, where the inaccuracies of the perturbative expansions are more pronounced.
\end{abstract}
\maketitle

\section{Introduction}
The thermodynamic behavior of a weakly interacting Bose gas is a topic of central interest in the physics of ultracold atoms and more generally in quantum statistical physics~\cite{book, Pethick-Smith}. A fundamental question concerns the effect of interactions, which modify the properties of the gas in a dramatic way even for very dilute systems. From an experimental point of view, mean-field and also higher-order effects in several thermodynamic quantities have been observed at very low temperatures and have been characterized as a function of the gas parameter $na^3$~\cite{PhysRevLett.107.135301}. In fact, at $T=0$, the properties of the interacting gas in the dilute regime are fully determined by the density $n$ and the $s$-wave scattering length $a$. At finite temperature, thermal effects are active even in the absence of interaction resulting in a weaker, and generally harder to measure, dependence of the thermodynamic quantities on interactions. The shift of the Bose-Einstein condensation (BEC) critical temperature with interactions has been carefully measured for trapped gases~\cite{PhysRevLett.106.250403} together with other quantities, such as the release energy, pointing out deviations with respect to the ideal gas predictions~\cite{PhysRevLett.77.4984}. More recently, the canonical equation of state of a homogeneous Bose gas was measured as a function of temperature in the condensed and normal phases showing clear evidence of effects of interaction in the pressure, chemical potential, and compressibility~\cite{PhysRevLett.125.150404}. Nevertheless, the observation of critical fluctuations and correlations beyond the mean-field description at finite temperature is still an open problem which requires more precise experimental schemes and also theoretical benchmarks more accurate than the available perturbative expansions~\cite{book, Pethick-Smith}.

In this paper we report precise unbiased path-integral Monte Carlo (PIMC) results of the thermodynamics of a dilute Bose gas as a function of temperature across the BEC transition. Careful extrapolation to the thermodynamic limit and focusing on deviations from the ideal gas behavior provide useful insights on the effects of interactions in the pressure and energy per particle at temperatures ranging from above the chemical potential scale in the condensed phase, to above the transition temperature in the normal phase.
Notably, we calculate the isothermal compressibility, finding indications of a  discontinuous behavior at the transition point, and we discuss the possibility that this is due to the finite data resolution, in analogy with the discontinuous specific heat in superconductors.
The temperature dependence of the contact parameter is also analyzed. The interaction strength in the gas is fixed at the value $na^3=10^{-6}$, which is known to be small enough for a universal description of the ground state in terms of the $s$-wave scattering length to be valid, without additional effects from the microscopic details of the interatomic potential~\cite{PhysRevA.60.5129}. At finite temperature the universal regime is expected to hold if $na^3\ll10^{-5}$~\cite{PhysRevA.69.053625}.
We also use our unbiased results to benchmark approximate theories valid to first and second order in the coupling constant. The first-order approach corresponds to the mean-field Hartree-Fock (HF) theory and its improvement is provided by the second-order Popov theory which includes the effects of quantum and thermal fluctuations beyond the mean-field description~\cite{PhysRevA.102.063303}. The comparison clearly indicates the extension of the critical region surrounding the transition point where approximate theories are known to fail. We notice that similar comparisons with PIMC results were already carried out in Refs.~\cite{PhysRevA.74.043621, Capogrosso_Sansone_2010}. Here we rely on more precise results for pressure and energy which allow for a detailed study of the effects of interactions.

The paper is organized as follows. In Sec.~\ref{sec:theo} we discuss the theoretical methods used in this study. In particular, in Sec.~\ref{subsec:pimc} we briefly outline the PIMC method based on the pair-product approximation and we explain how we calculate the relevant physical observables. In Sec.~\ref{subsec:hf} we present the basic ingredients of the Hartree-Fock and Popov perturbation schemes emerging from the analysis of the free-energy expressions both below and above the transition temperature. Starting from these free-energy expressions, we derive in the Appendix all thermodynamic quantities used in the comparison with the outcomes of PIMC simulations.  All results are discussed in Sec.~\ref{sec:results} and finally in Sec.~\ref{sec:conclusions} we add some concluding remarks.

\section{Theoretical Methods}
\label{sec:theo}
\subsection{Path Integral Monte Carlo}
\label{subsec:pimc}

We consider a system of $N$ identical bosons with mass $m$ described by the following Hamiltonian with two-body interactions:
\begin{equation}
H=- \frac{\hbar^2}{2m}\sum_{i=1}^N\nabla_i^2+\sum_{i<j}V(|{\bf r}_i-{\bf r}_j|) \;,
\label{hamiltonian}
\end{equation}
where ${\bf r}_i$ indicates the particle position vector. For the interatomic potential we use the hard-sphere (HS) model
\begin{equation}
V(r)=\left\{  \begin{array}{cc} +\infty & (r<a)  \\
                                   0    & (r>a)  \end{array} \;,\right.
\label{HS}
\end{equation}
where the hard-sphere diameter $a$ corresponds to the $s$-wave scattering length. In the limit where the gas parameter $na^3$ is small, the properties of the interacting quantum-degenerate gas do not depend on the details of the specific model of interatomic forces and are universal in terms of the scattering length $a$.

The partition function $Z$ of a bosonic system with inverse temperature
$\beta=(k_BT)^{-1}$, where $k_B$ is Boltzmann's constant, is defined as the trace over all states of the density
matrix $\rho=e^{-\beta H}$ properly symmetrized.
The partition function
satisfies the  convolution equation
\begin{eqnarray}
Z &=& \frac{1}{N!}\sum_P \int d{\bf R}\, \rho({\bf R},P{\bf R},\beta) =
\frac{1}{N!}\sum_P \int d{\bf R}
\label{PIMC1}\\ \nonumber
  &\times&  \int d{\bf R}_2 \cdots \int d{\bf R}_M \,\rho({\bf R},{\bf R}_2,\delta_{\tau})\cdots
  \rho({\bf R}_M,P{\bf R},\delta_{\tau}) \;,
\end{eqnarray}
where $\delta_{\tau}=\beta/M$, ${\bf R}$ collectively denotes the position vectors
${\bf R}=({\bf r}_1,{\bf r}_2,\ldots,{\bf r}_N)$, $P{\bf R}$ denotes the
position vectors with permuted labels  $P{\bf R}=({\bf r}_{P(1)},{\bf
r}_{P(2)},\ldots,{\bf r}_{P(N)})$ and the sum extends over the $N!$
permutations of $N$  particles. The calculation of the partition function
in Eq.~\eqref{PIMC1} can be mapped to a classical-like simulation of polymeric
chains with the number of beads $M$ equal to the number of terms of the convolution
integral. In a PIMC calculation, one makes use of
suitable approximations for the density matrix  $\rho({\bf R},{\bf
R}^\prime,\delta_{\tau})$ at the higher temperature $1/\delta_{\tau}$ in Eq.~\eqref{PIMC1}
and performs the  multidimensional integration over ${\bf R},{\bf
R}_2,\ldots,{\bf R}_M$ as well as the sum over permutations $P$ by  Monte
Carlo sampling~\cite{PhysRevB.30.2555, PhysRevB.36.8343, RevModPhys.67.279}.
In particular, the sampling over permutations can be performed in a very efficient way by using the worm algorithm~\cite{PhysRevLett.96.070601},
where both closed and open polymers are simulated at the same time.

In the case of dilute systems, featuring a large ratio of length scales between the average interparticle distance and the range of interactions, a particularly convenient approximation scheme for the high temperature density matrix, which significantly reduces the number $M$ of beads needed for convergence, is the pair-product ansatz~\cite{RevModPhys.67.279}
\begin{equation}
\rho({\bf R},{\bf R}^\prime,\delta_{\tau})=\prod_{i=1}^N\rho_1({\bf r}_i,{\bf r}_
i^\prime,\delta_{\tau})\prod_{i<j}
\frac{\rho_{rel}({\bf r}_{ij},{\bf r}_{ij}^\prime,\delta_{\tau})}
{\rho_{rel}^0({\bf r}_{ij},{\bf r}_{ij}^\prime,\delta_{\tau})} \;.
\label{PIMC4}
\end{equation}
In the above equation $\rho_1$ is the single-particle ideal-gas density matrix
\begin{equation}
\rho_1({\bf r}_i,{\bf r}_i^\prime,\delta_{\tau})=\left(\frac{m}{2\pi\hbar^2\delta_{\tau}}
\right)^{3/2}
e^{-({\bf r}_i-{\bf r}_i^\prime)^2m/(2\hbar^2\delta_{\tau})} \;,
\label{PIMC5}
\end{equation}
and $\rho_{rel}$ is the two-body density matrix of the
interacting system, which depends on
the relative coordinates  ${\bf r}_{ij}={\bf r}_i-{\bf r}_j$ and ${\bf
r}_{ij}^\prime={\bf r}_i^\prime-{\bf r}_j^\prime$,
divided by the corresponding ideal-gas term
\begin{equation}
\rho_{rel}^0({\bf r}_{ij},{\bf r}_{ij}^\prime,\delta_{\tau})=
\left(\frac{m}{4\pi\hbar^2\delta_{\tau}}\right)^{3/2}
e^{-({\bf r}_{ij}-{\bf r}_{ij}^\prime)^2 m/(4\hbar^2\delta_{\tau})} \;.
\label{PIMC6}
\end{equation}

The advantage of decomposition (\ref{PIMC4}) is that the two-body density matrix at the inverse temperature $\delta_{\tau}$,
$\rho_{rel}({\bf r},{\bf r}^\prime,\delta_{\tau})$, can be calculated exactly for a given potential $V(r)$, thereby solving by construction the two-body problem
which is the most relevant when the system is dilute. For the HS potential a simple and remarkably accurate analytical approximation of the high-energy two-body density matrix
is due to Cao and  Berne~\cite{doi:10.1063/1.463076}. The result is given by
\begin{eqnarray}
\frac{\rho_{rel}({\bf r},{\bf r}^\prime,\delta_{\tau})}
{\rho_{rel}^0({\bf r},{\bf r}^\prime,\delta_{\tau})}&=&
 1 -\frac{a(r+r^\prime)-a^2}{rr^\prime} \\ \nonumber
&\times& e^{-[rr^\prime +a^2-a(r+r^\prime)](1+\cos\theta)m/(2\hbar^2\delta_{\tau})} \;,
\label{CaoBerne}
\end{eqnarray}
where $\theta$ is the angle between the directions of $\bf{r}$ and $\bf{r}^\prime$.

In a PIMC simulation the statistical expectation value of a given operator $O({\bf R})$,
\begin{equation}
\langle O\rangle = \frac{1}{Z}\frac{1}{N!}\sum_P
\int d{\bf R}\, O({\bf R}) \rho({\bf R},P{\bf R},\beta) \;,
\label{PIMC2}
\end{equation}
is calculated by generating stochastically a set of configurations $\{{\bf
R}_i\}$, sampled from a probability density  proportional to the
symmetrized density matrix, and then by averaging over the set of values
$\{O({\bf R}_i)\}$. The results are exact for the equilibrium state of the microscopic Hamiltonian and are affected only by statistical uncertainty. In the present study we calculate internal energy $U$ and pressure $p$, whose thermodynamic estimators are defined by
\begin{eqnarray}
U&=&-\frac{1}{Z}\frac{dZ}{d\beta}
\nonumber\\
p&=&\frac{1}{\beta Z}\frac{dZ}{dV} \;.
\label{Eq:Thermo}
\end{eqnarray}
In addition, the above quantities can be calculated in PIMC simulations using the so called virial estimators~\cite{RevModPhys.67.279}, which usually suffer from smaller statistical fluctuations. Furthermore, we calculate the pair distribution function
\begin{equation}
g(r)=\frac{2}{nN}\Big{\langle}\sum_{i<j}\delta({\bf r}_i-{\bf r}_j-\bf{r})\Big{\rangle} \;.
\label{Eq:pair}
\end{equation}
From its short-range behavior one extracts the contact parameter
\begin{equation}
C=16\pi^2n^2\lim_{r\to0}\frac{r^2 a^2}{(r-a)^2}g(r) \;,
\label{Eq:contact}
\end{equation}
where the limiting procedure means distance scales much smaller than the average interparticle separation ($r\ll n^{-1/3}$) and much larger than the range of interactions ($r\gg a$).

We perform simulations in a cubic box of volume $V=L^3$ with periodic boundary conditions. Most simulations are performed in the canonical ensemble and, in terms of the density $n=\frac{N}{V}$, the gas parameter is kept fixed at the value $na^3=10^{-6}$ while the temperature $T$ is varied. The number of beads $M$ utilized ranges from $M=12$ to larger values in order to ensure full convergence at all temperatures.

Permutation sampling is performed using the worm algorithm~\cite{PhysRevLett.96.070601}. This is computationally more efficient  than the sampling algorithm described in Ref.~\cite{RevModPhys.67.279}, which was employed in a previous PIMC study of interacting Bose gases~\cite{PhysRevA.74.043621}. In particular, it significantly reduces the statistical uncertainty for large system sizes
~\footnote{For example, at $T=T_c^0$ we obtain $U/(Nk_BT_c^0)=0.7950(6)$ for $N=1024$, which is statistically consistent but significantly more precise than the corresponding  result  $U/(Nk_BT_c^0)=0.816(10)$ from Ref.~\cite{PhysRevA.74.043621}. It is worth mentioning that, in the algorithm of Ref.~\cite{RevModPhys.67.279}, the sampling of different winding-number sectors in large systems is exponentially suppressed. However, the possible bias induced in the equation of state is expected to vanish in the thermodynamic limit.}, allowing us to discern the residual finite-size effects and, therefore, to perform precise extrapolations to the thermodynamic limit. Specifically, the energy and pressure results for particle numbers from $N=128$ to $N=512$ are fitted with straight lines as a function of $1/N$.
 For selected temperatures, especially close to the critical point where finite size effects are more relevant, we extend the considered system sizes to $N=1024$ to check the reliability of the  thermodynamic-limit extrapolations.
Notably, the improved precision allows us to precisely identify the interaction effects and to extract the isothermal compressibility. Notice that the latter quantity was not determined before via unbiased computational techniques.
Furthermore, the worm algorithm allows us to perform grand-canonical simulations, and we employ them to extract the isothermal compressibility also from the fluctuation-dissipation relation.

\subsection{Hartree-Fock and Popov theories}
\label{subsec:hf}
Theories of dilute Bose gases have a long history and are the subject matter of various textbooks~\cite{Fetter-Walecka, Popov, Pethick-Smith, book, Svistunov-Prokofev}. At $T=0$ the situation is rather clear, mean-field terms are the leading ones in the small parameter $na^3$, and the so-called Lee-Huang-Yang beyond-mean-field contributions appear as higher-order corrections. At finite $T$, instead, on the one hand, the natural expansion parameter involves the condensate density $n_0$ rather than the total density, and on the other hand, the condensate is depleted not only by interactions but also by purely thermal effects. This complicates things, and various schemes are possible. The simplest one is the Hartree-Fock approximation, which is based on single-particle excitations and accounts for the leading-order contributions arising from interaction at temperatures sufficiently higher than the chemical potential~\cite{Popov}. In principle, the HF scheme is self-consistent and therefore includes terms to all orders in the coupling constant. Since, however, such terms beyond the leading ones are incorrect, we refer to the HF theory as the contribution from only the first-order mean-field terms~\cite{PhysRevA.102.063303}. The effects of quantum and thermal correlations beyond the mean-field regime are properly accounted for in the Popov scheme, based instead, on Bogoliubov excitations ~\cite{Popov, Capogrosso_Sansone_2010, PhysRevA.102.063303}. Also this scheme is self-consistent, but contributions beyond the next-to-leading order in the parameter $\Lambda$ (see below) are not reliable, and we neglect them in our analysis~\cite{Capogrosso_Sansone_2010, PhysRevA.102.063303}. The resulting approach is referred to as second-order Popov theory.

The structure of the second-order Popov scheme is provided with some details in Ref.~\cite{PhysRevA.102.063303}. Here we just report the result for the Helmholtz free energy in the condensed phase, $T<T_c$, where $T_c$ labels the BEC transition temperature. The energy density reads
\begin{eqnarray}\label{Eq:FPOP1}
\frac{F}{V} &=& \frac{g}{2} (n^2 + {n_T^0}^2) + \frac{k_BT}{V} \sum_\mathbf{k} \ln \left(1 - e^{-\beta E_\mathbf{k}}\right)
\nonumber\\
&+& \frac{16\sqrt{2}}{15\sqrt{\pi}} \left(\frac{m}{2\pi \hbar^2}\right)^{3/2} \Lambda^{5/2} \, ,
\end{eqnarray}
in terms of the Bogoliubov quasi particle spectrum $E_\mathbf{k}=\sqrt{\epsilon_\mathbf{k}^2+2\Lambda\epsilon_\mathbf{k}}$, where $\epsilon_\mathbf{k}=\frac{\hbar^2k^2}{2m}$ is the free-particle kinetic energy, and of the effective chemical potential calculated to lowest order $\Lambda=g(n-n_T^0)$. Here $g=\frac{4\pi\hbar^2a}{m}$ is the usual coupling constant fixed by the $s$-wave scattering length $a$ and $n_T^0=\frac{\zeta(3/2)}{\lambda_T^3}$, with $\zeta(x)$ being the Riemann zeta function and $\lambda_T=\sqrt{\frac{2\pi\hbar^2}{mk_BT}}$ being the thermal wave length, is the density of thermally excited atoms in a non-interacting gas. By performing the integration over momenta in Eq.~\eqref{Eq:FPOP1} one finds the more compact result
\begin{eqnarray}\label{Eq:FPOP2}
\frac{F}{V} &=& \frac{g}{2} (n^2 + {n_T^0}^2) + \frac{16\sqrt{2}}{15\sqrt{\pi}} \left(\frac{m}{2\pi \hbar^2}\right)^{3/2} \Lambda^{5/2}
\\
&-& \frac{4\tau}{3\sqrt{\pi}} \left(\frac{m}{2\pi \hbar^2}\right)^{3/2} \Lambda^{5/2}
\int_0^\infty \frac{dx}{e^x-1}\left(u-1\right)^{3/2} \, ,
\nonumber
\end{eqnarray}
\begin{figure}[t]
    \begin{center}
        \includegraphics[width=8.5cm]{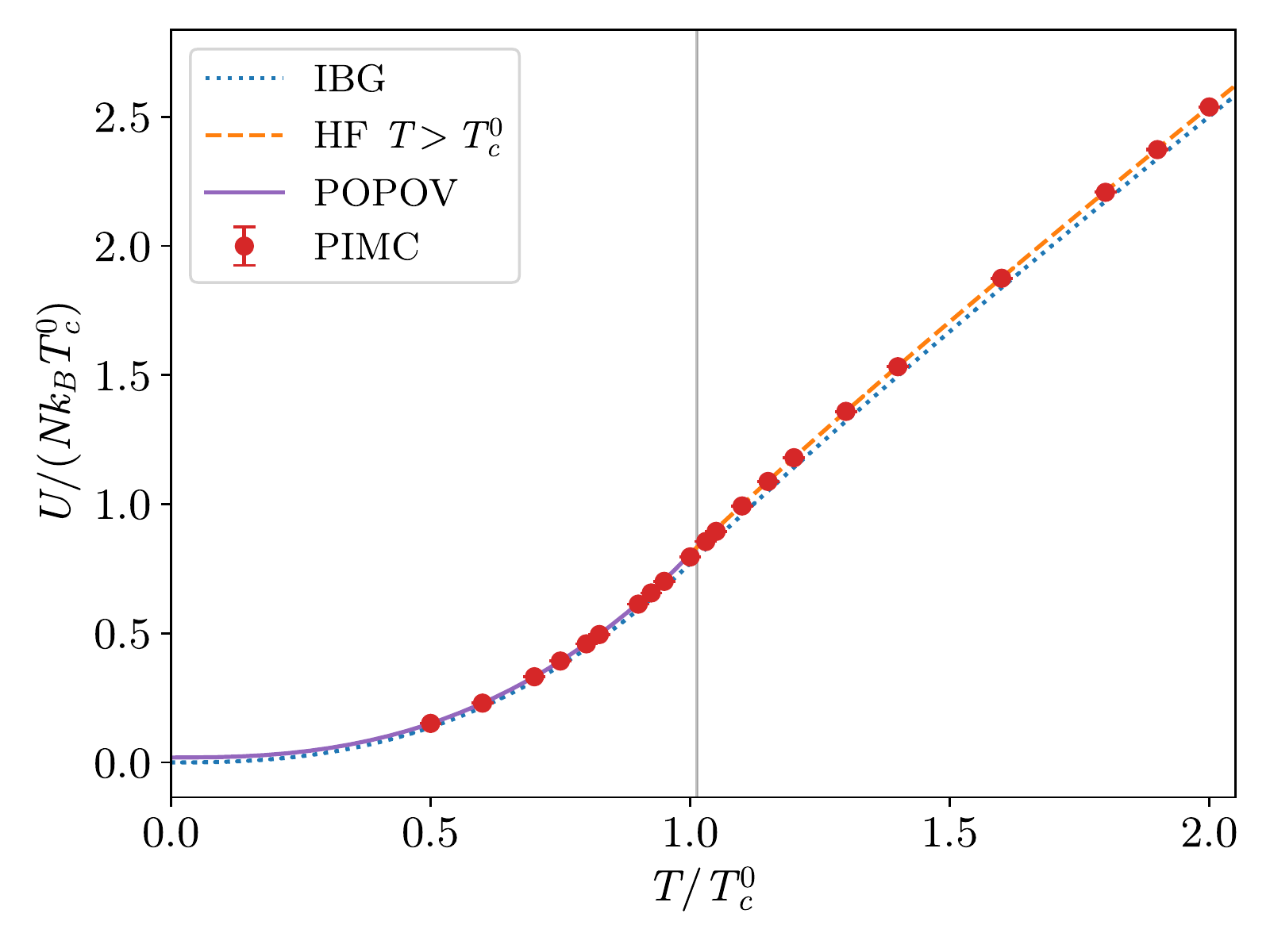}
        \caption{Internal energy per particle as a function of temperature. The lines refer to the ideal Bose gas result in Eq.~\eqref{U0}, the HF theory above $T_c^0$ in Eq.~\eqref{Eq:Unormal} and the result from Popov theory in Eq.~\eqref{Eq:UPOP}. The vertical line refers to the transition point $T_c$ slightly shifted from $T_c^0$ due to interaction effects [see Eq.~\eqref{Eq:Tc}].}
        \label{fig1}
    \end{center}
\end{figure}
where $\tau=\frac{k_BT}{\Lambda}$ is a reduced temperature and we define $u=\sqrt{1+(\tau x)^2}$. As discussed in Ref.~\cite{PhysRevA.102.063303}, the above result accounts correctly for quantum and thermal fluctuations up to order $\Lambda^{5/2}$ including the Lee-Huang-Yang correction to the ground-state energy. This is provided by the second term in the first line of the equation above which does not vanish as $T\to0$. Starting from Eq.~\eqref{Eq:FPOP2} one can easily neglect beyond mean-field corrections and reduce the free energy to the HF form holding to linear order in the coupling constant $g$:
\begin{eqnarray}
\frac{F_{\text{HF}}}{V} &=& \frac{g}{2} (n^2 - {n_T^0}^2)+gnn_T^0-\frac{\zeta(5/2)}{\zeta(3/2)}k_BTn_T^0 \, .
\label{Eq:FHF}
\end{eqnarray}
Above the transition point, i.e., in the normal phase $T>T_c$, the free-energy density can be obtained from perturbation theory valid to linear order in $g$. This approach corresponds again to HF theory and yields a constant interaction shift added to the free energy of a non-interacting gas
\begin{equation}
\frac{F}{V}=gn^2-\frac{k_BT}{\lambda_T^3}g_{5/2}(z) + n\, k_B T \ln z\;.
\label{Eq:Fnormal}
\end{equation}
Here, $z$ is an effective fugacity which determines the total density of the gas via the equation $n\lambda_T^3=g_{3/2}(z)$. Furthermore, $g_\nu(z)$ stand for the usual special Bose functions. We point out that the above interaction shift can also be derived from the $s$-wave contribution to the second coefficient of the virial expansion in the limit $a\ll\lambda_T$. Higher-order contributions to the virial coefficients arising from interactions are expected to depend on the details of the interatomic potential and are therefore no longer universal in the $s$-wave scattering length~\cite{Pathria}. One should also notice that the expressions written above in Eqs.~(\ref{Eq:FPOP1}) and (\ref{Eq:Fnormal}) are continuous at the transition point of the ideal gas
\begin{equation}
k_BT_c^0=\frac{2\pi\hbar^2}{m}\left(\frac{n}{\zeta(3/2)}\right)^{2/3} \;,
\label{Eq:Tc0}
\end{equation}
where $n_T^0=n$ and $\Lambda=0$.

\section{Results}
\label{sec:results}

\begin{figure}
    \begin{center}
        \includegraphics[width=8.5cm]{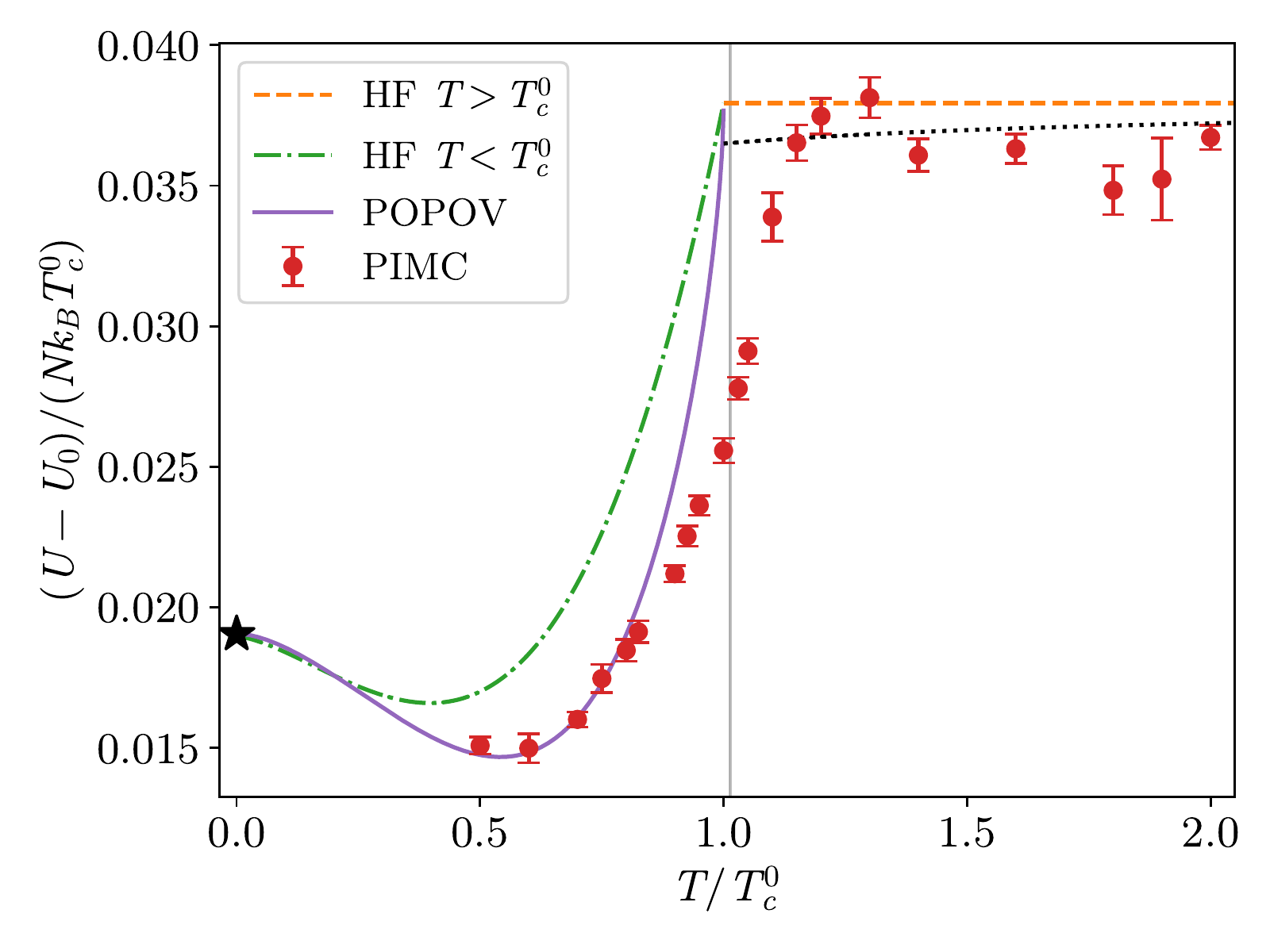}
        \caption{Energy shift $U-U_0$ per particle as a function of temperature. The lines refer to the HF theory above $T_c^0$ from Eq.~\eqref{Eq:Unormal}, the first-order HF theory from Eq.~\eqref{Eq:UHF} and the result from second-order Popov theory in Eq.~\eqref{Eq:UPOP}. The dotted line at $T/T_c^0>1$ includes the contribution from the third virial coefficient (see text). At $T=0$ we also report the ground-state energy from Ref.~\cite{PhysRevA.60.5129} obtained using the diffusion Monte Carlo method (star). Notice that for this result the error bar is smaller than the size of the symbol. The vertical line is as in Fig.~\ref{fig1}.}
        \label{fig2}
    \end{center}
\end{figure}

The results reported in this section in the canonical ensemble all refer to a specific value of the gas parameter, namely	 $\bar{n}a^3=10^{-6}$. We start by discussing our findings for the energy and the pressure. In Fig.~\ref{fig1} we show the internal energy per particle $U/N$ as a function of the rescaled temperature $T/T_c^0$. The comparison with the ideal Bose gas (IBG) result:
\begin{equation}
\frac{U_0}{N}=\left\{  \begin{array}{cc}  \frac{3}{2}\frac{k_BT}{n\lambda_T^3}\zeta(5/2) & (T<T_c^0) \,,  \\
    \\
                                \frac{3}{2}\frac{k_BT}{n\lambda_T^3}g_{5/2}(z)   & (T>T_c^0) \,,  \end{array} \right.
\label{U0}
\end{equation}
indicates the presence of interaction effects even though on the scale of the Fig.~\ref{fig1} they are hardly visible. More useful information can be extracted instead from Fig.~\ref{fig2}, where we plot the energy difference $\delta U=U-U_0$, directly pointing out deviations from the IBG law. PIMC results are compared with Hartree-Fock theory above the transition temperature and with both Hartree-Fock and Popov theories in the condensed phase (see the Appendix).

First we notice a visible discrepancy between first-order HF and second-order Popov results, pointing out the relevance of beyond mean-field effects in the region of intermediate temperatures compared to the small Lee-Huang-Yang correction in the ground-state energy. We also observe that the second-order Popov theory nicely matches exact quantum Monte Carlo results at $T=0$, where we report the energy obtained  with the diffusion Monte Carlo method in Ref.~\cite{PhysRevA.60.5129}, up to temperatures close to but sizably below the critical point. In fact, as shown in Fig.~\ref{fig2}, in the range $0.8 T_c^0\lesssim T\lesssim 1.2T_c^0$ any mean-field-based approach fails to accurately describe the system which appears to be dominated by critical fluctuations. The fact that the critical region is significantly wide even for very small values of the interaction parameter was noted in Ref.~\cite{PhysRevA.69.053625}, where an estimate of the size of the critical region yields $|\Delta T|/T_c\sim\alpha n^{1/3}a$, with $\alpha$ being a numerical coefficient of order $100$.

\begin{figure}
    \begin{center}
        \includegraphics[width=8.5cm]{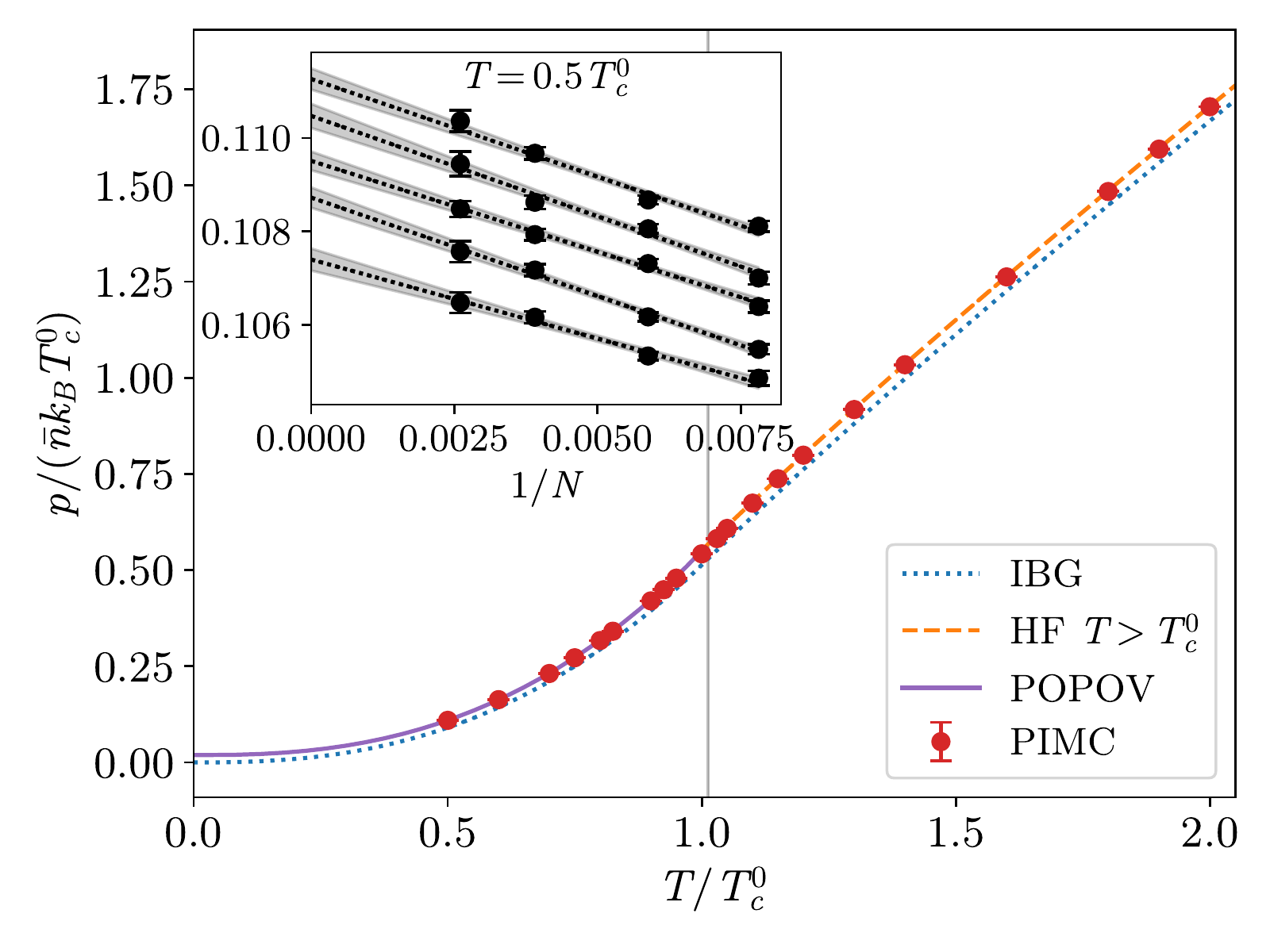}
        \caption{Pressure as a function of temperature. As in Fig.~\ref{fig1}, the lines refer to the ideal Bose gas result, the HF theory above $T_c^0$ in Eq.~\eqref{Eq:pnormal} and the result from Popov theory in Eq.~\eqref{Eq:pPOP}. Inset: Finite-size scaling as a function of $1/N$ at fixed temperature $T=0.5T_c^0$ for slightly different densities around the central value $\bar{n}$ in terms of which $T_c^0$ is determined. Starting from the uppermost one, lines correspond to $1.05, 1.025, 1.00, 0.975, 0.95$ of the reference density $\bar{n}$. The vertical line is as in Fig.~\ref{fig1}.}
        \label{fig3}
    \end{center}
\end{figure}

\begin{figure}
\begin{center}
\includegraphics[width=8.5cm]{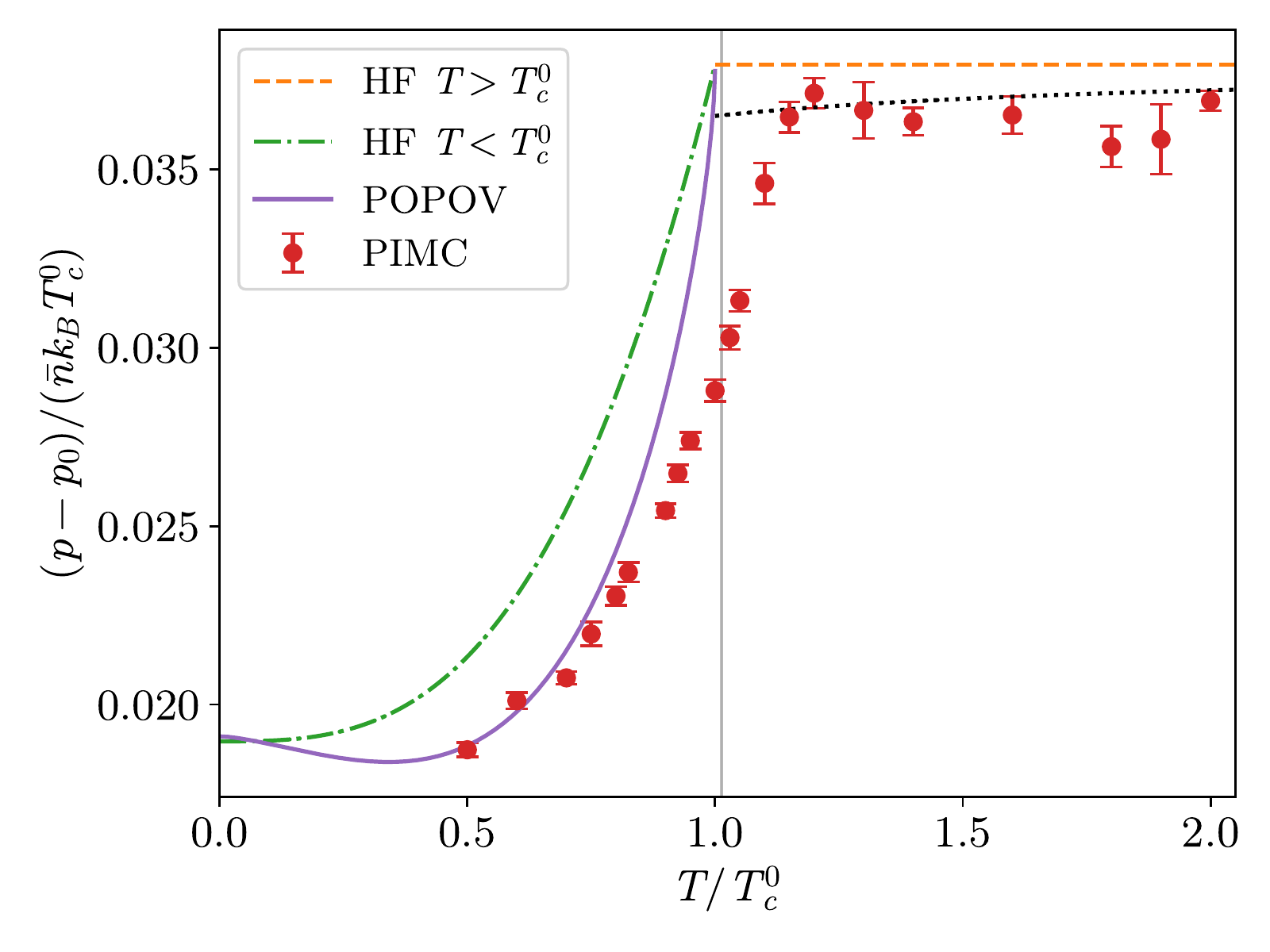}
\caption{Pressure shift $p-p_0$ as a function of temperature. As in Fig.~\ref{fig2}, the lines refer to the HF theory above $T_c^0$ from Eq.~\eqref{Eq:pnormal}, the first-order HF theory from Eq.~\eqref{Eq:pHF} and the result from second-order Popov theory in Eq.~\eqref{Eq:pPOP}. The vertical line is as in Fig.~\ref{fig1}. The dotted line at $T/T_c^0>1$ includes the contribution from the third virial coefficient (see text).}
\label{fig4}
\end{center}
\end{figure}

In the high-temperature regime the interaction shift from HF theory is constant in $T$ and is in good agreement with PIMC results. We emphasize that in this high temperature regime interactions provide a much smaller relative correction to the energy, resulting in the larger error bars shown in Fig.~\ref{fig2}. To analyze this region further we notice that higher-order contributions in $a/\lambda_T$ to the second virial coefficient, which depend on the specific model of interatomic potential, are completely negligible in the temperature range we consider~\cite{Pathria, PhysRev.116.250}. On the contrary, the leading contribution to the third virial coefficient depends only on $a$ and is proportional to $a^2/\lambda_T^2$~\cite{PhysRev.116.250}. The corresponding effect on the pressure is the small negative shift $\delta p=-nk_BT(n\lambda_T^3)^2 4\frac{a^2}{\lambda_T^2}$, yielding a similar shift in the internal energy $\delta U=-Nk_BT(n\lambda_T^3)^2 4\frac{a^2}{\lambda_T^2}$, which is shown in Fig.~\ref{fig2} with a dotted line. Our results, despite the large error bars, are in reasonable agreement with this higher-order correction. Similar results are obtained for the pressure $p$ and the pressure shift $p-p_0$, where $p_0=\frac{2}{3}\frac{U_0}{V}$ is the pressure of the non-interacting gas. These are shown in Figs.~\ref{fig3} and~\ref{fig4}.

A crucial ingredient to reliably extract the results for energy and pressure in the thermodynamic limit is the extrapolation to infinite system sizes. An example of the procedure followed is shown for the pressure in the inset of Fig.~\ref{fig3}, where at a fixed temperature we report the values of $p$ obtained for different system sizes $N$. If $N$ is large enough a linear fit in $1/N$ is used to extract $p$ in the thermodynamic limit. In Fig.~\ref{fig3} this procedure is repeated for various values of the density around the central density $\bar{n}$. This allows us to calculate the derivative $dp/dn$ and therefore the isothermal compressibility $\kappa_T$ using the relation
\begin{equation}
\frac{1}{\kappa_T}=n\frac{\partial p}{\partial n} \;.
\label{Eq:compress}
\end{equation}

\begin{figure}
    \begin{center}
        \includegraphics[width=8.5cm]{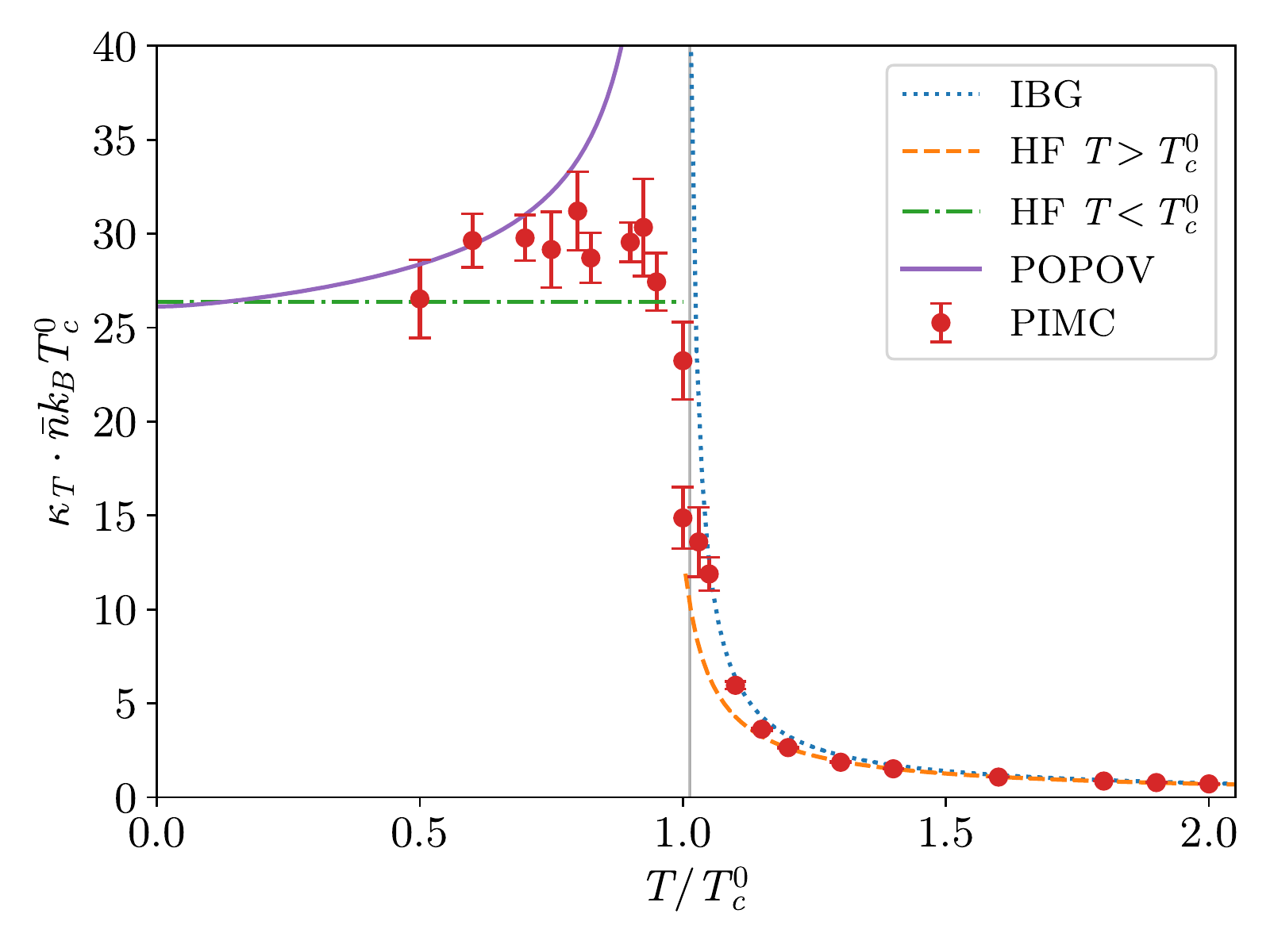}
        \caption{Isothermal compressibility $\kappa_T$ in units of $1/(\bar{n}k_BT_c^0)$ as a function of temperature. Above $T_c^0$ the lines refer to the ideal Bose gas and HF theory from Eq.~\eqref{Eq:compress_normal}. Below $T_c^0$, they refer to the first-order HF theory from Eq.~\eqref{Eq:kHF} and the second-order Popov theory from Eq.~\eqref{Eq:kPOP}. The vertical line is as in Fig.~\ref{fig1}.}
        \label{fig5}
    \end{center}
\end{figure}

\begin{figure}
    \begin{center}
        \includegraphics[width=8.5cm]{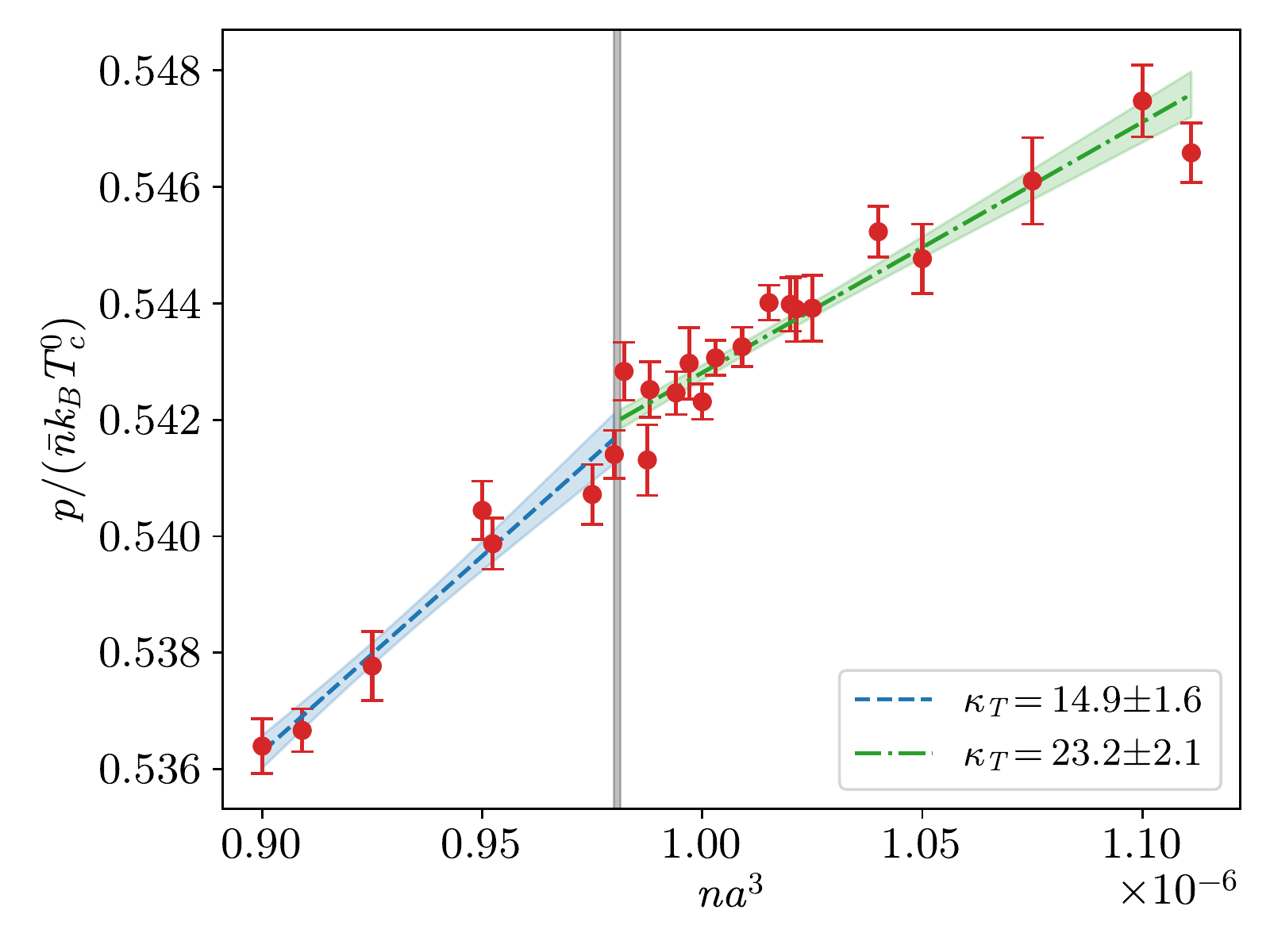}
        \caption{Gas pressure for slightly different densities around the central value $\bar{n}$ at the fixed temperature corresponding to $T_c^0$ for the density $\bar{n}$. The vertical line indicates the critical density $n_c$ from Eq.~\eqref{Eq:Tc}. The linear fits to the PIMC results provide the inverse compressibility below and above the transition point showing the discontinuity at $n_c$. The corresponding values of $\kappa_T$ are reported in the legend in units of $1/(\bar{n}k_BT_c^0)$. }
        \label{fig6}
    \end{center}
\end{figure}

The results for $\kappa_T$ are shown in Fig.~\ref{fig5}. Notice that the statistical uncertainty is larger for $\kappa_T$ due to the difficulty of extracting reliably small variations of pressure with respect to density. To this aim we consider at least four values of density around $\bar{n}$, ranging within a few percent of its value, and we use a linear or quadratic fit to extract $1/\kappa_T$. Above the transition point we find good agreement with the compressibility of the ideal Bose gas with small corrections due to interactions. Below the transition $\kappa_T$ is weakly dependent on temperature. In particular, critical fluctuations strongly suppress the divergent peak predicted by Popov theory, and beyond-mean-field effects rapidly become very small at low temperature, where $\kappa_T$ is consistent with the value from first-order HF theory. It is worth mentioning that a broad maximum in $\kappa_T$ below the transition point was observed in the experimental study of the equation of state of a unitary Fermi superfluid~\cite{Ku2012563}. The behavior at the critical point is best illustrated in Fig.~\ref{fig6}, where results for $p$ at fixed temperature are shown as a function of the density around the central value $\bar{n}$. A linear fit to these data yields $\kappa_T^{-1}$, which shows an apparent discontinuity at the critical density $n_c$. Furthermore, the value of $n_c$ is consistent with the predicted interaction shift~\cite{PhysRevLett.87.120402, PhysRevLett.87.120401, PhysRevA.65.013606}
\begin{equation}
T_c=T_c^0[1+c(an^{1/3})] \;,
\label{Eq:Tc}
\end{equation}
where $c=1.29\pm0.05$. The extracted values of the compressibility above ($n<n_c$) and below ($n>n_c$) the critical temperature are reported in the legend in Fig.~\ref{fig6}, and they are also shown in Fig.~\ref{fig5} as the two values assigned to the critical temperature $T_c$.

To further inspect the apparent discontinuous behavior of the compressibility $\kappa_T$ at the transition point, we perform PIMC simulations in the grand-canonical ensemble.
Here $\kappa_T$ can be determined from the following fluctuation-dissipation relation:
\begin{equation}
\label{eqFD}
\kappa_T = \frac{\left<n^2\right> - \left<n\right>^2 }{\left<n\right>^2} \frac{V}{k_B T},
\end{equation}
where $\left<n\right>$ and $\left<n^2\right>$ indicate grand-canonical expectation values at the chosen chemical potential $\mu$.

\begin{figure}
    \begin{center}
        \includegraphics[width=8.5cm]{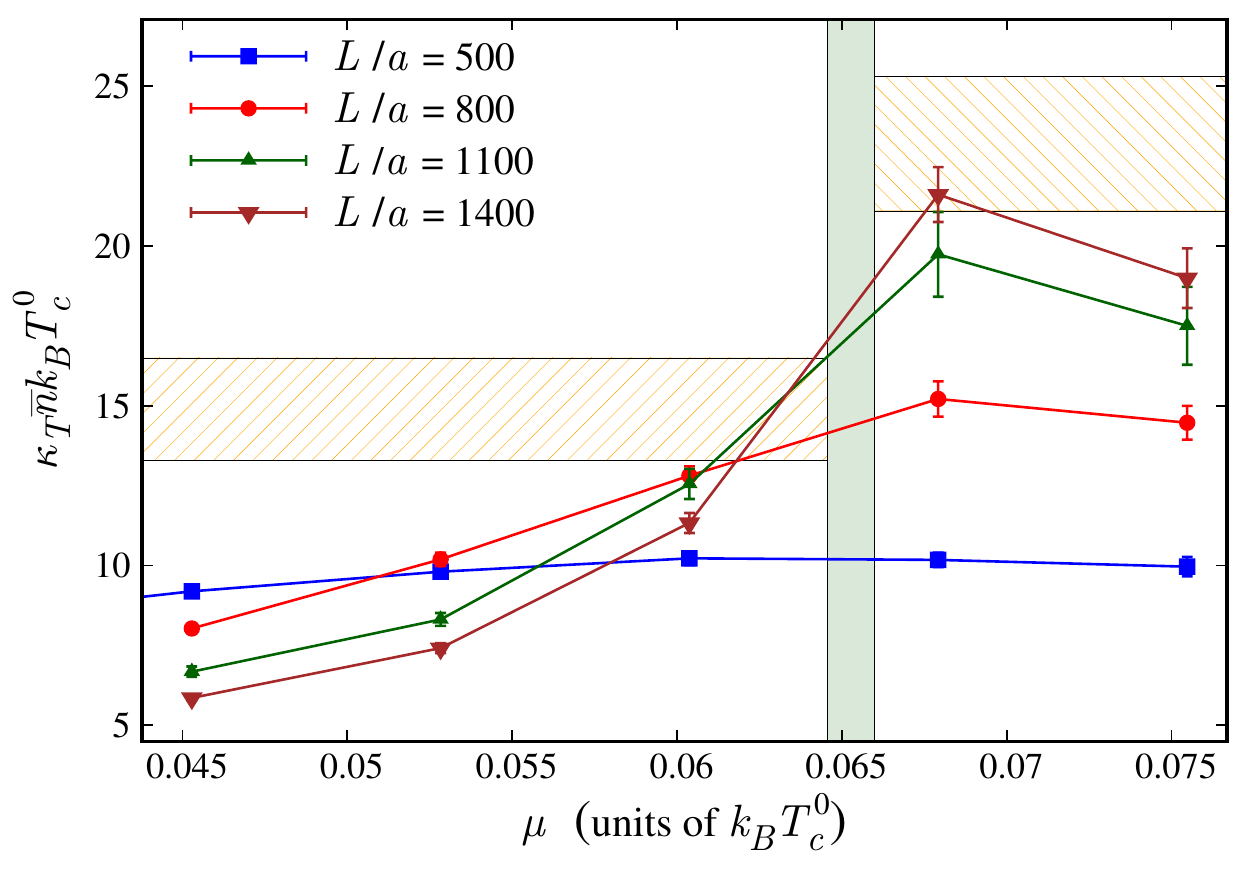}
        \caption{Isothermal compressibility in the grand-canonical ensemble as a function of the chemical potential and for different system sizes $L/a$. Results are obtained via the fluctuation-dissipation relation (\ref{eqFD}) at the fixed temperature $T_c^0$, corresponding to the ideal gas critical temperature for the density $\bar{n}$. The values of $\bar{n}$ and $T_c^0$ are also used to normalize the compressibility $\kappa_T$. The vertical band indicates the critical chemical potential $\mu_c$, including the estimated uncertainty due to finite-size effects (see text). The horizontal bands refer instead to the estimates of the compressibility above and below the critical point obtained in the canonical ensemble.}
        \label{fig7}
    \end{center}
\end{figure}

Figure~\ref{fig7} displays PIMC results obtained at the fixed temperature $T_c^0$ corresponding to the density $\bar{n}$, for different box sizes up to $L/a=1400$. It is worth noticing that this box size corresponds to average particle numbers up to $N\approx 3500$ for the largest chemical potentials we consider.
The value of $\mu$ is varied across the BEC transition. The critical chemical potential is identified as the one for which the density expectation value is $\left< n \right> = n_c$, and the critical density $n_c$ is computed assuming the weak coupling formula~(\ref{Eq:Tc}). Cubic fitting functions are used to describe the dependence of the average density on $\mu$.

It should be mentioned that the so-determined critical chemical potential slightly varies with the inverse box size $1/V$, and we perform a linear extrapolation (excluding the smallest box size) to estimate the thermodynamic-limit result $\mu_c$ and the corresponding uncertainty.

We find that $\kappa_T$ suddenly grows when $\mu$ is increased beyond the critical point $\mu_c$, and that this growth sharpens as the box size increases. The compressibility values obtained just before and beyond $\mu_c$ are consistent with the corresponding canonical results at the same density, in the normal and in the condensed phases, respectively. These findings are consistent with the indications of a discontinuous behavior of $\kappa_T$ at the BEC transition obtained from the canonical simulations, as discussed above.
However, we should point out that, due to the inevitably finite resolution, our data cannot unambiguously discern a discontinuity from a very sharp, but continuous, behavior.
This scenario is similar to the one observed in superconducting materials, where  discontinuities in specific heats are routinely employed to characterize the superconducting transition (see, e.g., Ref.~\cite{JUNOD2000214}).
Indeed, in weakly correlated superconductors the specific heat displays a well defined jump, as predicted by the BCS theory. Instead, in strongly correlated materials the specific heat is continuous, as in liquid $^4$He~\cite{PhysRevLett.76.944}, suggesting that with sufficient resolution one would always observe a continuous behavior. Indeed, such systems belong to the universality class of the three-dimensional $XY$ model, corresponding to a continuous specific heat.
However, considering the critical scaling relation from the renormalization group theory, the continuity results from a relation between amplitudes which, to the best of our knowledge, has not been demonstrated to be universal~\cite{PhysRevB.68.174518,PhysRevLett.51.2291}. This means that for other systems in the same universality class, such as the weakly interacting Bose gas, discontinuities are not formally ruled out.

\begin{figure}
    \begin{center}
        \includegraphics[width=8.5cm]{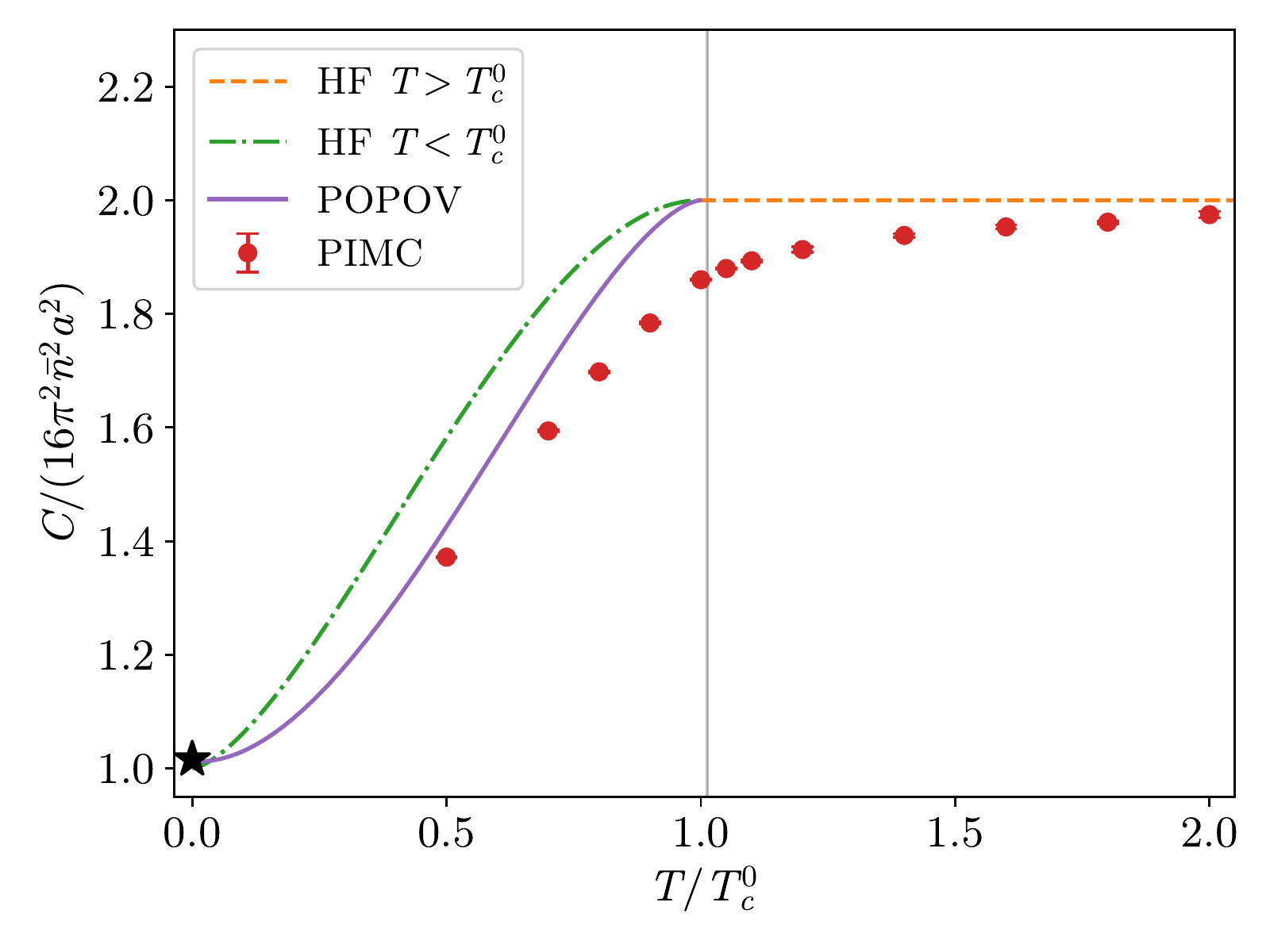}
        \caption{Contact parameter $C$ in units of $16\pi^2 \bar{n}^2a^2$ as a function of temperature. The lines refer to the virial expansion in Eq.~\eqref{Eq:Cnorm}, the first-order Hartree-Fock theory from Eq.~\eqref{Eq:CHF} and the second-order Popov theory of Eq.~\eqref{Eq:CPOP}. The value at $T=0$ (star symbol) is the result of a diffusion Monte Carlo calculation at the density $\bar{n}$. The vertical line is as in Fig.~\ref{fig1}.}
        \label{fig8}
    \end{center}
\end{figure}

Finally, we discuss the contact parameter defined through the derivative of the free energy density with respect to the inverse of the scattering length
\begin{equation}
C=-\frac{8\pi m}{\hbar^2}\frac{\partial F/V}{\partial 1/a} \;.
\label{Eq:contact1}
\end{equation}
This quantity can be conveniently calculated in a PIMC simulation from the short-range behavior of the pair correlation function [see Eq.~(\ref{Eq:contact})]. We point out that $C$ is sensitive only to interaction effects, both below and above the transition, and therefore represents a quantity better suited to compare to approximate theories than the previously considered energy, pressure, and compressibility. The results obtained from the short-range behavior of the pair-correlation function via Eq.~\eqref{Eq:contact} are shown in Fig.~\ref{fig8}. We observe that $C$ is in reasonable agreement with Popov theory at low temperatures, showing deviations in the vicinity of the transition point. Furthermore, we find that $C$ approaches the HF value at high temperatures above the transition. The experimental investigation of the contact parameter across the phase transition could provide a useful tool to gain insight into the role of critical fluctuations in a weakly interacting Bose gas~\cite{PhysRevLett.108.145305, doi:10.1126/science.aax5850}.

\section{Conclusions}
\label{sec:conclusions}

We have carried out a precise and careful study of interaction effects in the thermodynamic properties of a weakly interacting Bose gas using exact PIMC simulations. The use of the worm algorithm enabled us to gain sufficient precision in the calculation of energy, pressure, and compressibility to discriminate between approximate theories such as first-order Hartree-Fock and second-order Popov theory. In general, we observe good agreement with HF results at high temperature and with Popov results at low temperature. However, in a large window of intermediate temperatures including the transition point, such approximate schemes fail to reliably describe the thermodynamic behavior of the gas which is strongly affected by critical fluctuations not accounted for by mean-field based approaches. In particular, we provided a PIMC study of the contact parameter across the transition temperature, and we pointed out large deviations compared to the predictions of perturbation schemes. We also found indications of discontinuous behavior of the compressibility at the BEC transition and an estimate of the jump was provided. We hope that our study will stimulate further experimental efforts devoted to a quantitative analysis of the role played by interactions in shaping the thermodynamics of a dilute Bose gas.
All PIMC results presented in this paper are freely available from Ref.~\cite{zenodo}.

\section*{Acknowledgments} This work was supported by the Italian Ministry of University and Research under the PRIN2017 project CEnTraL 20172H2SC4. S.P. acknowledges PRACE for awarding access to the Fenix Infrastructure resources at Cineca, which are partially funded by the European Union’s Horizon 2020 research and innovation program through the ICEI project under the Grant Agreement No. 800858.
S. P. also acknowledges the Cineca award under the ISCRA initiative, for the availability of high performance computing resources and support. The authors would like to thank F.~Werner for fruitful discussions.

\appendix
\setcounter{equation}{0}
\renewcommand\theequation{A\arabic{equation}}
\section*{APPENDIX: Popov and Hartree-Fock results for the thermodynamic quantities}

In this Appendix we derive explicit formulas holding below and above the transition temperature $T_c$ for the thermodynamic quantities discussed in Sec.~III. In the condensed phase we use both the full second-order Popov result for the Helmholtz free energy in Eq.~\eqref{Eq:FPOP2} and the mean-field expression in Eq.~\eqref{Eq:FHF}, obtained from first-order Hartree-Fock theory.

\subsection{Pressure}

The pressure is defined as $p=-\frac{\partial F}{\partial V}$. In the condensed phase one gets the result
\begin{eqnarray}\label{Eq:pPOP}
p&=& \frac{g}{2} (n^2 - {n_T^0}^2) + \left(\frac{m\Lambda}{2\pi \hbar^2}\right)^{3/2} \Bigg[ \frac{8\sqrt{2}\Lambda}{5\sqrt{\pi}}
+ \frac{8\sqrt{2}gn_T^0}{3\sqrt{\pi}}
\nonumber\\
&+& \frac{2\tau\Lambda}{\sqrt{\pi}}\int_0^\infty \frac{dx}{e^x-1}\left(u-1\right)^{3/2} \left(\frac{2}{3}+\frac{1}{u}\right)
\\
&+& \frac{2\tau gn_T^0}{\sqrt{\pi}}\int_0^\infty \frac{dx}{e^x-1}\frac{(u-1)^{3/2}}{u} \Bigg] \;,
\nonumber
\end{eqnarray}
which one derives straightforwardly from Eq.~\eqref{Eq:FPOP2}. By neglecting beyond mean-field corrections, this result reduces to the Hartree-Fock expression valid to linear order in the coupling strength
\begin{equation}\label{Eq:pHF}
p_{\text{HF}}= \frac{g}{2}(n^2+{n_T^0}^2)+\frac{\zeta(5/2)}{\zeta(3/2)}k_BTn_T^0 \;.
\end{equation}
Above the transition point the expression for the pressure reads
\begin{equation}\label{Eq:pnormal}
p=gn^2+\frac{k_BT}{\lambda_T^3}g_{5/2}(z) \;.
\end{equation}
This result corresponds to the pressure of a non-interacting gas shifted by the constant term $gn^2$. Notice that both the HF and Popov results are continuous at $T=T_c^0$.

\subsection{Energy density}

The energy density is defined as $\frac{U}{V}=\frac{F}{V}+\frac{TS}{V}$, where $S=-\frac{\partial F}{\partial T}$ is the entropy of the gas. The result for $U/V$ is given by
\begin{eqnarray}\label{Eq:UPOP}
\frac{U}{V}&=& \frac{gn^2}{2} -g{n_T^0}^2 + \left(\frac{m\Lambda}{2\pi \hbar^2}\right)^{3/2} \Bigg[ \frac{16\sqrt{2}\Lambda}{15\sqrt{\pi}}
+\frac{4\sqrt{2}gn_T^0}{\sqrt{\pi}}
\nonumber\\
&+& \frac{2\tau\Lambda}{\sqrt{\pi}} \int_0^\infty \frac{dx}{e^x-1}\left(u-1\right)^{3/2} \frac{u+1}{u}
\\
&+& \frac{3\tau gn_T^0}{\sqrt{\pi}}\int_0^\infty \frac{dx}{e^x-1}\frac{(u-1)^{3/2}}{u}\Bigg] \;.
\nonumber
\end{eqnarray}
The corresponding Hartree-Fock expression reads
\begin{equation}\label{Eq:UHF}
\frac{U_{\text{HF}}}{V}=\frac{g}{2}n^2+g{n_T^0}^2-\frac{g}{2}nn_T^0+\frac{3}{2}\frac{\zeta(5/2)}{\zeta(3/2)}k_BTn_T^0 \;.
\end{equation}
Similarly to the pressure, the energy density in the normal phase is given by the ideal-gas value shifted by $gn^2$,
\begin{equation}\label{Eq:Unormal}
\frac{U}{V}=gn^2+\frac{3}{2}\frac{k_BT}{\lambda_T^3}g_{5/2}(z) \;.
\end{equation}
Both HF and Popov expressions for the internal energy are continuous at $T=T_c^0$.

\subsection{Compressibility}

The inverse isothermal compressibility is defined in Eq.~(\ref{Eq:compress}) and it requires the calculation of the derivative of the pressure with respect to the density. A direct calculation using Eq.~\eqref{Eq:pPOP} yields the result
\begin{eqnarray}\label{Eq:kPOP}
\frac{1}{\kappa_T} &=&gn^2\left[1+\frac{4\sqrt{2}}{\sqrt{\pi}}\left(\frac{m}{2\pi \hbar^2}\right)^{3/2}g\Lambda^{1/2} - \left(\frac{m}{2\pi \hbar^2}\right)^{3/2} \right.
\nonumber \\
&\times&  \left. g\Lambda^{1/2}\frac{\tau}{\sqrt{\pi}}\int_0^\infty \frac{dx}{e^x-1}\frac{(u-1)^{3/2}}{u}\frac{3u+2}{u^2} \right] \;.
\end{eqnarray}
The first-order result reduces to just the first term in the above equation
\begin{equation}\label{Eq:kHF}
\left(\frac{1}{\kappa_T}\right)_{\text{HF}}=gn^2 \;,
\end{equation}
whereas above the transition point one finds
\begin{equation}
\frac{1}{\kappa_T}=2gn^2+n^2\lambda_T^3\frac{k_BT}{g_{1/2}(z)} \;.
\label{Eq:compress_normal}
\end{equation}
We notice that expression (\ref{Eq:kPOP}) diverges at $T=T_c^0$, whereas, according to mean-field theory, the inverse compressibility exhibits a discontinuity from $gn^2$ below the transition to $2gn^2$ just above $T_c^0$.\\

\subsection{Contact parameter}

The contact parameter is defined in Eq.~\eqref{Eq:contact1}. Using Eq.~\eqref{Eq:FPOP2} for the free energy one finds the result
\begin{eqnarray}
  C &=& 16\pi^2n^2a^2 \left\lbrace
    1+\left(\frac{T}{T_c^0}\right)^3 +
  \right. \nonumber \\
  &+& \frac{1}{\zeta(3/2)} \Big(\frac{\Lambda}{k_BT_c^0}\Big)^{3/2}  \left[1 - \Big(\frac{T}{T_c^0}\Big)^{3/2}\right] \times\\
 &~& \times  \left.\left(\frac{16\sqrt{2}}{3\sqrt{\pi}}+\frac{4\tau}{\sqrt{\pi}}\int_0^\infty \frac{dx}{e^x-1}\frac{(u-1)^{3/2}}{u}\right)\right\rbrace \,. \nonumber
\label{Eq:CPOP}
\end{eqnarray}
Within the mean-field approximation only the first two terms survive, yielding
\begin{equation}
C_{\text{HF}}=16\pi^2n^2a^2\left[1+2\left(\frac{T}{T_c^0}\right)^{3/2} -\left(\frac{T}{T_c^0}\right)^3\right] \;.
\label{Eq:CHF}
\end{equation}
Finally, above the transition temperature one finds the temperature independent result
\begin{equation}
C=32\pi^2n^2a^2 \;.
\label{Eq:Cnorm}
\end{equation}
We notice that, according to HF theory, $C$ is continuous at $T_c^0$, whereas the Popov expression yields $C=64\pi^2n^2a^2$ at $T=T_c^0$, therefore predicting a discontinuous jump at the transition.

\bibliography{BosePIMC}

\end{document}